\documentclass{article}

\usepackage{amsmath,amssymb,amsfonts}
\usepackage{graphicx}
\usepackage[letterpaper, margin=1in]{geometry}
\usepackage{url}
\usepackage{authblk}
\DeclareMathOperator*{\argmax}{argmax}

\usepackage[backend=bibtex,style=numeric]{biblatex} 
\title{On Usefulness of the Deep-Learning-Based Bug Localization Models to Practitioners}
\author[1]{Sravya Polisetty}
\author[2]{Andriy Miranskyy}
\author[3]{Ayse Bener}
\affil[1,2,3]{Ryerson University, Toronto, Canada}
\affil[ ]{\{sravya.polisetty, avm, ayse.bener\}@ryerson.ca}
\date{}

\begin{document}
\maketitle

\begin{abstract}

\textit{Background}: Developers spend a significant amount of time and efforts to localize bugs. In the literature, many researchers proposed state-of-the-art bug localization models  to help developers localize bugs easily. The practitioners, on the other hand, expect a bug localization tool to meet certain criteria, such as trustworthiness, scalability, and efficiency. The current models are not capable of meeting these criteria, making it harder to adopt these models in practice.  Recently, deep-learning-based bug localization models have been proposed in the literature, which show a better performance than the state-of-the-art models. 

\textit{Aim}: In this research, we would like to investigate whether deep learning models meet the expectations of practitioners or not. 

\textit{Method}: We constructed a Convolution Neural Network and a Simple Logistic model to examine their effectiveness in localizing bugs. We train these models on five open source projects written in Java and compare their performance with the performance of other state-of-the-art models trained on these datasets. 

\textit{Results}: Our experiments show that although the deep learning models perform better than classic machine learning models, they meet the adoption criteria set by the practitioners only partially.

\textit{Conclusions}: This work provides evidence that the practitioners should be cautious while using the current state of the art models for production-level use-cases. It also highlights the need for standardization of performance benchmarks to ensure that bug localization models are assessed equitably and realistically.

\end{abstract}

\section{Introduction}\label{sec:Intro}
Software quality assurance is the process that ensures that the software being developed, meets all the expected quality standards \cite{Sommerville:2010:SE:1841764}. However, software systems are often shipped with bugs. A software \textit{bug} is an anomaly in the software product that causes the software to perform incorrectly or to behave in an unexpected way \cite{braude2016software}. Large software projects contain large number of bugs that are encountered by users.  For instance, the Eclipse \cite{EclipseP14:online} project had nearly 13,016 bugs reported in a span of one year (2004 - 2005), with an average of 37 bugs reported per day, and a maximum of 220 bugs reported in a single day \cite{Anvik:2006:ABR:1134285.1134457}.

On detecting a bug or fault in the software, a user or a software tester, reports the bug in a document called a \textit{bug report}  and logs it in a bug tracking system, such as Bugzilla \cite{Bugzilla32:online} or JIRA \cite{JiraIssu97:online}. Before a bug report is considered valid, it goes through various stages of quality checks. During these stages, a bug is checked for duplicity, validity, and completeness. After a bug passes through these stages, it is assigned to an expert (``Assignee"), who refers to the information in the bug report to identify the source files which need to be modified, in order to solve the issue in the bug report. Bug reports typically consist of various fields (such as Title, Version, Component, Product, Quality Assurance Contact, Assignee, Reported Date, Importance, and Attachments), which contain description of the bug, screen shots or snapshots of an error message, stack traces, etc. 

After a bug is assigned to a developer, he/she has to find the root cause of the bug in the source code and then fix the root cause. This process is typically manual and can take 30\%-40\% of the total time needed to fix a problem~\cite{Murtaza:2014:ESU:2588914.2589302}.  This is a painstaking process, especially for large software projects with hundreds of thousands of source files. As a result, the bug fixing time increases, along with maintenance cost of the project. Although the bug fixing task constitutes sub-tasks like understanding the bug, validating the bug, locating the cause of the bug, and finally fixing the bug, it is the \textit{``locating the cause of the bug  (Bug Localization)"}, that consumes most of the developer's time \cite{chang2007simulation}. Hence the need for automated tools or approaches which perform bug localization.

Many researchers have proposed various automated approaches or models, to pinpoint the locations of bugs in the source files. Though all these state-of-the-art models are able to localize bugs to some extent, none of them could meet the expectations of the software practitioners, who are the end users of these models \cite{Kochhar:2016:PEA:2931037.2931051}. These models are unable to bridge the lexical gap between the bug reports and source code. To bridge this gap, recently, researchers \cite{Huo-2016, Lam-2015} have proposed deep-learning-based bug localization models.

Deep Learning \cite{Deeplear99:online} is becoming increasingly popular. This is because deep-learning-based models have proved to perform better than traditional machine learning (ML) models in the areas of image processing \cite{krizhevsky2012imagenet}, speech recognition \cite{hinton2012deep}, and natural language processing (NLP) \cite{Collobert:2008:UAN:1390156.1390177}. This is why many recent works apply deep learning to software data to solve various software engineering problems, such as  defect prediction \cite{yang2015deep}, user profiling~\cite{Curro:2017:BUP:3102962.3102976}, software artifact traceability \cite{Guo:2017:SES:3097368.3097370}, code suggestion \cite{White:2015:TDL:2820518.2820559}, processing programming languages \cite{Mou:2016:CNN:3015812.3016002}, and bug localization \cite{Huo-2016, Lam-2015}. In the past few years, different architectures of Deep Neural Nets (DNN) have been proposed to localize bugs. The motivation behind our study is based on the fact that these deep-learning-based models have shown to outperform the state-of-the-art traditional ML-based bug localization models \cite{ Huo-2016, Lam-2015, Lam-2017, Huo-2017}. Hence, our \textit{primary objective} is to examine the effectiveness of these models, in meeting the expectations of the software practitioner. Our \textit{secondary objectives} are (a) to compare the performance of our deep learning model with a traditional ML model, (b) to train deep-learning-based bug localization models independently on five open source software datasets and compare the performance of these models on each of the datasets, and (c) to observe the effect of varying source files on the performance of the model. 

Many DNNs have been introduced in the literature; thus, we can leverage previously reported statistics for many of them. We also implement a particular Convolutional Neural Network, deemed CNN, that is based on a popular network architecture~\cite{kim2014convolutional} to run custom tests. We reach our objectives by answering the following research questions:

\textbf{RQ1}: How can we minimize the lexical gap between natural language texts in bug reports and technical/domain corpus in source code files in order to automatically localize bugs?

\textbf{RQ2}: How effective are the state-of-the-art traditional and the recent deep-learning-based models, in meeting the expectations of the software practitioner?

\textbf{RQ3}: How do the CNN models perform compared to baseline  ML models, such as Simple Logistic model, on software bug localization data?

\textbf{RQ4}: How do the CNN models perform when trained on different open source software bug localization datasets?

\textbf{RQ5}: How does varying the source files in the dataset affect the performance of the CNN and Simple Logistic Models?

In order to address RQ1, we give a detailed explanation of the existing state-of-the-art traditional and deep learning models and why a deep-learning-based model like CNN could be a potential solution for bridging the lexical gap between bug reports and source files. For addressing RQ2, we train CNN on open source datasets, which have been widely used in the past bug localization studies. We then compare the performance of the CNN models and other state-of-the-art traditional and deep learning models, with the expectations of the software practitioner \cite{Kochhar:2016:PEA:2931037.2931051}. 
In order to address RQ3, we train a Simple Logistic model on our datasets. As bug localization is a Learning-To-Rank Information Retrieval (IR) problem, we need calibrated results (those which do not have ties between the scores assigned to the suggested source files for a bug report) in order to be able to correctly measure the performance of the resultant IR model. Therefore, we need a logistic regression model, like the Simple Logistic model, that can give us calibrated output scores and can also meet the computation and memory constraints, which are inherent when dealing with large datasets like the ones used in this study. Next, for answering RQ4, we train the CNN model on five open source bug localization datasets and evaluate the performance of the model on each of them. We address the next research question, RQ5 by varying the buggy source files in each dataset. The major contributions of this study can be summarized as follows:
\begin{itemize}

    \item We examine the relevance of a deep-learning-based bug localization model to the software industry.
    
    \item We evaluate a deep learning model (CNN) against a traditional ML model (Simple Logistic) on bug localization data\footnote{We chose ``vanilla'' CNN and logistic regression models as simple baselines, since (based on parsimony principle) we prefer simple models to complex ones~\cite{bishop2012pattern}. The former model has shown good results for sentence classification in general~\cite{kim2014convolutional} and traceability in particular~\cite{Huo-2017}. The latter model is one of the simplest classic binary classifiers. We created them to 1) measure required timing and amount of resources and 2) assess the effect of the search space size on the performance of these models, which helped us in answering RQs.}.
    
    \item We extract the source files for all the bug reports based on the bug-commit mapping (reported in a widely used dataset \cite{Ye2014}) for five open source datasets and train bug localization models on all of them. We publicly share our dataset via figshare~\cite{our_data}.
    
    \item We empirically show the effect of varying the source files on the performance of both the CNN and Simple Logistic bug localization models.

\end{itemize}

\section{Background and Related Literature}\label{sec:Background}
We now provide an overview of a survey \cite{Kochhar:2016:PEA:2931037.2931051}, which has been one of the motivating factors of this study. Kochhar et al. \cite{Kochhar:2016:PEA:2931037.2931051} surveyed 386 practitioners from more than 30 countries across 5 continents, about their expectations of research in bug localization. They then compared what practitioners need, and the current state of research, by performing a literature review of papers on bug localization techniques. The survey found that even the best performing studies could not satisfy even 75\% of the respondents in the survey. Many of the studies that can satisfy 50\% of the practitioners, work at a granularity that is considered very coarse by most of the practitioners, i.e., class- or file-level. The survey revealed that almost all the practitioners are willing to adopt a bug localization technique, if it satisfies the below criteria:

\begin{itemize}

\item \textit{Preferred level of granularity}: $\approx$52\% of practitioners in the survey prefer method-level granularity. Only $\approx$26\% of practitioners prefer file-level granularity.

\item \textit{Minimum success criteria}: About $\approx$91\% of developers gave Top-5 (or better) as the minimum success criteria, i.e., an artifact of interest should be reported in the Top-5 records returned by the localization technique. Only $\approx$17\% of the practitioners would be satisfied with Top-10 (or better).

\item \textit{Trustworthiness}: In order to achieve a satisfaction rate of 50\%, 75\%, and 90\%, a bug localization model has to be successful 50\%, 75\%, and 90\% of the time respectively.

\item \textit{Scalability}: In order to achieve a developer satisfaction rate of 50\%, 75\%, and 90\%, the bug localization model needs to be scalable enough to deal with programs of size 10,000, 100,000, and 1,000,000 Lines Of Code (LOC), respectively.

\item \textit{Efficiency}: In order to achieve a developer satisfaction rate of at least 50\%, the model should have a run time of less than a minute. This threshold for efficiency was approved by almost 90\% of the practitioners who participated in the survey.

\end{itemize}

We now discuss some significant research in bug localization, that uses the traditional IR and ML approach. We also briefly mention research related to the application of deep learning models to different software engineering problems.

\subsection{Traditional Approaches}
The automated approaches in bug localization can be broadly classified into two categories: \textit{dynamic} and \textit{static} approaches. The semantics of the program and its execution information with test cases, i.e., pass/fail execution traces are used in the dynamic approach. These approaches can be categorized into: \textit{Spectrum-based fault localization} and \textit{Model-based fault localization}. The Spectrum-based approach uses program traces to correlate program elements at statement-, block-, function-, or component-level in a program to the program failures. Saha et al. \cite{saha2011fault} have proposed a model that uses this approach for localizing faults in data-centric programs, which interact with databases. Model-based fault localization techniques have been proposed by Feldman et al. \cite{feldman2006two} and Mayer et al. \cite{mayer2007model}. These models are based on expensive logical reasoning of formal models of programs but are often more accurate than the Spectral-based models. 

The second type of automated approach, i.e., the static approach rely only on the source code and the bug reports information. These static approaches can be further classified into two groups: \textit{program analysis-based} and \textit{IR-based}. The program analysis-based approach uses predefined bug patterns to localize bugs. Hovemeyer et al. \cite{Hovemeyer} used this approach to propose a model called \textit{FindBugs}. This model does not perform well as it gives a large number of false positives and also false negatives \cite{Thung}. The second type of static approach uses IR or  ML techniques like TF-IDF, LSA, LDA, VSM, rVSM, and Naive Bayes \cite{khoury2016advances}. These approaches treat the bug localization problem as a \textit{Learning-To-Rank} IR problem. 

Rao et al. \cite{Rao-2011} probed into many of these IR techniques and concluded that simpler techniques, such as TF-IDF, work better than complex ones. Lukins et al. \cite{Lukins} proposed a bug localization model using LDA, which is a topic-modelling approach. Zhou et al. \cite{ZhouZhang} proposed the rVSM approach, which is a refined vector space model to leverage the similarity between the bug reports and source files. Saha et al. \cite{Saha} proposed a structured retrieval model that employs the structure of bug reports and source code files, to achieve better performance. Kim et al. \cite{Kim} have also experimented using Naive Bayes algorithm, with the previously fixed files as classification labels. Ye et al. \cite{Ye:2014:LRR:2635868.2635874} used the adaptive Learning-To-Rank approach, to train the features extracted from source files, API (Application Programming Interface) descriptions, bug-fixing, and change history. Most of these state-of-the-art models consider the source code and bug reports to be of the same lexical space and try to correlate these artifacts by measuring their similarity in the same space. None of these traditional approaches for bug localization meet all the adoption criteria set by the practitioners \cite{Kochhar:2016:PEA:2931037.2931051}. 

\subsection{Deep Learning for NLP and Software Engineering}
Kim \cite{kim2014convolutional} has experimented on a basic Convolution Neural Network (CNN) on open source datasets, like Movie Reviews and proved that a simple CNN (with minimal hyper-parameter tuning while using static word vectors), is capable of performing exceptionally well on the sentence classification task. Kim's experiments proved that learning task-specific word vectors instead of static vectors improves the performance of the model. Zhang et al. \cite{DBLP:journals/corr/ZhangL15,DBLP:journals/corr/ZhangZL15} experimented on character-level CNN models for classifying text. Their experiments proved that CNN perform better than other traditional models (such as bag-of-words, n-grams, and TF-IDF models) and also other deep learning models (such as word-based CNN and Recurrent Neural Networks (RNN)). They also proved that CNNs do not require knowledge of words in terms of semantics or context of words, to perform text classification. Zhang et al. \cite{zhang2015sensitivity} performed a sensitivity analysis on the effect of varying hyper-parameters in a CNN model. This model proves to be very useful to researchers aiming to train CNN models on different datasets. Nguyen et al. \cite{Nguyen-2011} built a CNN model for relation extraction and classification tasks. 

Recently, deep learning models are being widely used to solve software engineering problems. Mou et al. \cite{Mou:2016:CNN:3015812.3016002} have proposed a novel tree-based CNN architecture (TBCNN) for processing programming languages. Curro et al. \cite{Curro:2017:BUP:3102962.3102976} proposed an automatic approach using Deep Convolutional Neural Network (DCNN) to extract information about user actions from publicly available video tutorials of a product. Guo et al. \cite{Guo:2017:SES:3097368.3097370} applied RNN and their variants to establish links between software engineering artifacts, namely Requirements and Design documents. Lam et al. \cite{Lam-2015} applied Restricted Boltzmann Machine (RBM)-based \cite{Restrict43:online} DNN in combination of rVSM on bug localization datasets. Huo et al. \cite{Huo-2016} adopted a \textit{Pairwise} Learning-To-Rank approach to classify the combined corpora of bug reports and source files into linked and non-linked records. They proposed a new architecture called Natural Language And Programming Language CNN (NP-CNN) which outperformed the other state-of-the-art bug localization models. They also proposed yet another new architecture using a combination of LSTM and CNN called LS-CNN. They compared the performance of this model with other state-of-the-art bug localization models and deep-learning-based models. Their experiments proved that the LS-CNN model outperforms all the models on the same datasets.

\section{Methodology}\label{sec: Methodology}
This section describes some of the concepts pertaining to the Pairwise Learning-To-Rank bug localization problem, deep learning, CNNs, Simple Logistic models. We also discuss the evaluation metrics used to measure the performance of the models in our study.

\subsection{The Pairwise Learning-To-Rank Problem}
The automated bug localization approaches use IR to identify the source files which have the potential fix for a given bug report. The query of such an IR model is a bug report and the retrieved documents is a ranked list of source files, with the most relevant source file at the top and the least relevant at the bottom of the list. The Pairwise approach for the Learning-To-Rank problem is approximated by a \textit{classification} problem. The text of each bug report  and source file  are merged into a single sentence and fed into a classification model. If a given bug was fixed by the change in a given file, then it is labeled as \textit{linked}, otherwise -- as \textit{non-linked}.  The Learning-To-Rank problem is now converted into a binary classification problem, with linked and non-linked class labels.

\subsection{Deep Learning}
Modern deep learning models and training methods originated from research in Artificial Neural Networks (ANNs). ANNs are inspired from biological neural networks that constitute the brain \cite{van2018artificial}. A DNN is an ANN with multiple hidden layers between the input and output layers \cite{bengio2009learning,schmidhuber2015deep}. The benefit of this complex structure of a DNN, is the ability to represent the data in multiple layers. This is one of the key advantages of DNNs over traditional machine learning models, in which humans have to explicitly extract features from the data, before training the model. ANNs automatically discover ``good internal representations", i.e., features that make the learning easier and more accurate, through \textit{backpropagation}  \cite{rumelhart1986learning}. The extra layers in a DNN enable composition of features from lower layers, potentially modeling complex data with fewer units than a similarly performing shallow network \cite{bengio2009learning}. They can model complex non-linear relationships and generate compositional models where an object is expressed as a layered composition of primitives \cite{szegedy2013deep}. In case of text, the object is a sentence encoded using Word2Vec \cite{mikolov2013efficient}, GloVe word embeddings, or One-Hot Encoding \cite{OnehotEn40:online}. In \textit{One-Hot Encoding}, each word is converted into a sparse representation with only one element of the vector set to 1, the rest being 0. For each token, its index in the vocabulary dictionary defines the position of the one-hot element in the resultant vector. For natural language data, Word2Vec or GloVe encoding proves better than One-Hot encoding, however different representations perform better for different tasks \cite{zhang2015sensitivity}. For natural language sentence classification, One-Hot encoding performs poorly compared to the other two encodings. However this may not be the case if the training data are very large \cite{zhang2015sensitivity}. Nevertheless, prior work \cite{kim2014convolutional} on sentence classification using CNN showed that Word2Vec gives the best performance compared to the other forms of encoding. We first experimented with Word2Vec encoding on our data. The resultant models performed poorly, which could be due to the fact that our corpus has more domain-related words/tokens than natural language words or tokens. We then experimented on One-Hot encoding of the corpus, which proved to be function better than the Word2Vec encoded models. 

\subsection{Convolution Neural Nets (CNNs)}
In this section we discuss basic concepts of CNNs. CNNs are responsible for a major breakthrough in the field of Image Classification \cite{krizhevsky2012imagenet}. A CNN consists of an input and an output layer, as well as multiple hidden layers. The hidden layers of a CNN typically consist of convolutional layers, pooling layers, and fully connected layers \cite{Convolut99:online}. The \textit{convolution} operation can be visualized as a sliding window function applied to a matrix. CNNs consist of several layers of convolutions with non-linear activation functions applied to the convolution results. Each region of the input is connected to a neuron in the output. These local connections are a result of convolutions applied to the input to compute the output. Each layer in a CNN consists of hundreds or even thousands of filters of varied sizes. The pooling layers subsample the input from the convolution layer.

\textit{Local Invariance} is another important aspect of CNN which means that the information about the exact occurrence of a feature is lost. This occurs because of the max pooling operation. \textit{Compositionality} is yet another important feature of a CNN. As explained earlier, CNNs are feedforward neural networks, where each
neuron in a layer receives input from the neurons in the previous layer. These local receptive fields, allow CNN to recognize more and more complex patterns in a hierarchical way, by combining lower-level, elementary features into higher-level features. 

\begin{figure}
\includegraphics[width=\columnwidth]{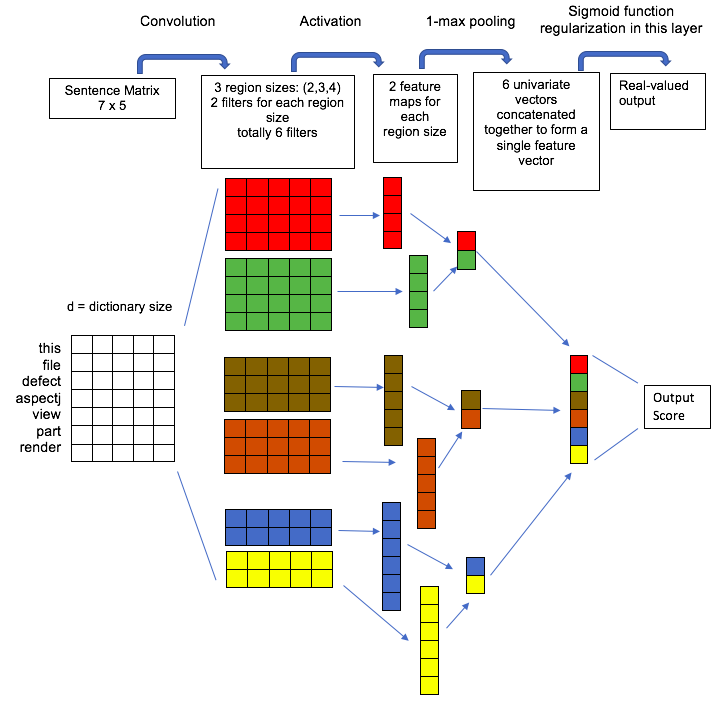}
\caption[Architecture of the CNN model] {Architecture of the CNN Model \cite{zhang2015sensitivity}}
\label{fig::1}
\end{figure}

CNNs perform well on textual data for many text-related tasks like sentiment analysis, text classification, question answering, and machine translation, due to the Local Invariance and Compositionality features. In text, n-grams can be constructed from lower-order features and ordering is crucial locally but not at the document or record level. For example, in trying to classify a movie review as positive or negative, we need to only capture local order, e.g., the fact that ``not" precedes ''bad" and so forth. 

We now discuss the architecture of the CNN model used in our study. Figure \ref{fig::1} illustrates the CNN model used in our study. The architecture of this model has been proposed  by Kim \cite{kim2014convolutional}. In Figure \ref{fig::1}, $d$ is the length of the dictionary, i.e., the number of unique words/tokens in the entire corpus of our dataset. Hence the dimensions of the input sentence matrix will be equal to $n \times d$, where $n$ is the maximum sentence length and $d$ is the dictionary size. There are three filter regions of size 2, 3, and 4, each of which have 2 filters. Each filter performs convolution on the input sentence matrix and generates feature maps of variable length, i.e., 2 feature maps for each region size. After this step, 1-max pooling is applied to each of these feature maps, which results in 6 univariate vectors. These vectors are then combined to form a single feature vector which is then fed into a sigmoid layer. The sigmoid layer uses the feature vector to produce a real-valued score which gives the degree of likeliness that the documents belongs to the positive class. For a given bug report, we rank its corresponding source files based on this score.

\subsection{Simple Logistic}
 
We now discuss the Simple Logistic model used in our study. This model was proposed by Landwehr et al. \cite{Landwehr2005}. It is a standalone logistic regression model with in-built attribute selection. Landwehr et al. built these regression models  using the \textit{LogitBoost} \cite{friedman2000additive} algorithm which selects a subset of attributes from the data. The pseudo code for the LogitBoost algorithm \cite{friedman2000additive} for $J$ classes is given below:
\begin{enumerate}
\item Start with weights $w_{ij} = 1/n$, $i = 1,\ldots,n$, $j = 1,\ldots,J$, $F_{j}(x) = 0$ and $p_{j}(x) = 1/J$  $\forall j$.

\item Repeat for $m = 1 \ldots M$:
    \begin{enumerate}
        \item Repeat for $j = 1 \ldots J$:
        \begin{enumerate}
        
            \item Compute working responses and weights in the $j^{th}$ class
            \begin{equation}
            z_{ij} = \frac{y^{\ast}_{ij} - p_{j}(x_{i})}{p_j(x_i)(1 - p_j(x_i))},
            \end{equation}
            \begin{equation}
            w_{ij} = p_j(x_i)(1 - p_j(x_i)).
            \end{equation}

            \item Fit the function $f_{mj}(x)$ by a weighted least-squares regression of $z_{ij}$ to $x_{i}$ with weights $w_{ij}$.
        \end{enumerate}
    
    \item Set $f_{mj}(x) \leftarrow \frac{J-1}{J}(f_{mj}(x) - \frac{1}{J} \sum_{k=1}^{J}f_{mk}(x))$, $F_{j}(x) \leftarrow F_{j}(x) + f_{mj}(x)$.
     
    \item Update 
    \begin{equation}
        p_{j}(x) = \frac{e^{F_{j}(x)}}{\sum_{k=1}^{J}e^{F_{k}(x)}}.
    \end{equation}
    \end{enumerate}
    
\item Output the classifier $\argmax_j F_{j}(x)$.
\end{enumerate}

$y_{i}$ is the class label of example $x_{i}$, $y^{\ast}_{ij}$ gives the observed class membership probabilities for instance $x_{i}$ and is defined as: 
\begin{equation}
    y_{ij}^{*}=
    \begin{cases}
    1, & \text{if $y_i = j$}\\
    0, & \text{if $y_i \neq j$}
    \end{cases}.
\end{equation}

The $p_j(x)$ are the
estimates of the class probabilities for an instance $x$ given by the model
fit so far. LogitBoost performs forward stage-wise fitting and tries to improve the model by adding a function $f_{mj}$ to the
committee $F_{j}$, fit to the response by weighted least-squared error. If we constrain $f_{mj}$ to be a linear function of only the attribute that results in a least squared error, then we arrive at the Simple Logistic algorithm that performs automatic attribute selection.

\subsection{Evaluation Metrics}

In this section we discuss the evaluation metrics used to measure the performance of the models in our study. From hereon, we will use the terms \textit{relevant} file and \textit{buggy} file interchangeably, as by relevant file we mean the file that has to be fixed to address a given bug report. Similarly, the terms \textit{non-relevant} and \textit{non-buggy} are used interchangeably, as by non-relevant we mean the source files which do not have to be altered to fix a given bug report.

\paragraph{Mean Average Precision ($MAP$)}
$MAP$ is computed as follows: 
\begin{equation}\label{eq:3.9}
     MAP = \sum_{i=1}^{n_{2}}
\sum_{j=1}^{n_{1}} \frac{Prec(j) * t(j)}{N_{i}},
\end{equation}
where $n_{1}$, $n_{2}$ are the number of candidate source files and retrieved bug reports, respectively, and $N_{i}$ is the number of relevant files to a bug report $i$, and $t(j)$ indicates whether the instance in rank $j$ is relevant or not (buggy or non-buggy). $Prec(j)$ is the precision at the given cut-off rank $j$ defined as
\begin{equation}\label{eq:3.10}
    Prec(j) = \frac{Q(j)}{j}, 
\end{equation}
where $Q(j)$ is the number of relevant source files in the top $j$ positions.

\paragraph{Mean Reciprocal Rank ($MRR$)} 
The reciprocal rank of a query response is the multiplicative inverse of its rank of the first correctly retrieved document: $1$ for the first place, $1/2$ for the second place, $1/3$ for the third place and so on. $MRR$ is the average of the reciprocal ranks of results for a sample of queries $Q$ and is given by:
\begin{equation}\label{eq:3.11}
MRR = \frac{1}{|Q|} \sum_{i=1}^{|Q|}\frac{1}{rank_{i}}.
\end{equation}

\paragraph{Top-$k$ Rank} It is the number of bugs whose relevant source files are ranked in the top $k$ of the returned results. Given a bug report, if the top  query results contain at least one source file that is relevant (buggy) to the bug report, then the particular bug is considered to be located. The percentage of all such located bugs gives the Top-$k$ Rank. 

\paragraph{Area Under Curve ($AUC$)} $AUC$ is the area under the Receiver Operating Characteristic Curve (ROC), which is a graphical plot that illustrates the diagnostic ability of a binary classifier system as its discrimination threshold is varied. $AUC$ values lie between $0.5$ and $1$, where $0.5$ denotes a bad classifier and $1$ denotes an excellent classifier. 

\paragraph{Cross Entropy Loss}
Cross entropy loss measures the performance of a classifier which outputs a probability score in the range $[0, 1]$. For binary classification, it is given by:
\begin{equation}
   L =  -[y\log(p)+(1-y)\log(1-p)],
\end{equation}
where $y$ is a binary indicator ($0$ or $1$) which gives the actual class label of a sample and $p$ is the model's predicted probability that a sample belongs to a class.

\section{Evaluation}\label{sec:Evaluation}
In this section we discuss the steps followed to prepare the datasets and the experiments performed, along with a brief discussion of the results obtained in our study.

\subsection{Data Extraction} 
We performed our experiments on the five open source benchmark datasets. The bug reports and bug-commit mappings are  publicly available via figshare~\cite{Ye2014}. Using these mappings, we extracted the actual source files from GitHub \cite{hubmanua43:online} repository of each project. We iterate over all the commits associated with fixing a bug; and for each commit, we checkout the before-fix version of the source code. The end result of the above extraction process will be the bug reports and their corresponding before-fix version of each file needed to fix a bug. For extracting the content of all the other source files (which were never altered to fix a bug), we clone the latest git repository of the project and extract all the Java source files in it. We identify the previously fixed files from the entire list of source files and label them as \textit{buggy}. The rest of the source files are labelled as \textit{non-buggy}. We share the resulting dataset via figshare~\cite{our_data}.

\subsection{Data Analysis}\label{sec:dat_analysis}
Table \ref{tab::1} shows the statistics for the datasets under study. In order to address RQ5, we perform all our experiments on three variations of source files. They are as follows:

\textit{All Files}: We consider all the source files and all the linked records in the traceability matrix.

\textit{Buggy Files}: We consider only the source files which have at least one bug report linked to it in the traceability matrix. 

\textit{Very Buggy Files}: We consider only the set of source files which have more than one bug linked to it in the traceability matrix.

\begin{table*}[t]
\caption{Bug reports and source files statistics}\label{tab::1}
\begin{center}
\resizebox{\textwidth}{!}{%
\begin{tabular}{|l|r|r|r|r|r|r|r|r|r|}
\hline
Dataset & Bug Reports & \multicolumn{3}{c|}{Source Files} & \multicolumn{2}{c|}{Linked Pairs}                                       & \multicolumn{3}{c|}{Non-linked Pairs}                                                         \\ \hline
        &  & All Files & Buggy Files & Very Buggy Files & \multicolumn{1}{c|}{All / Buggy} & \multicolumn{1}{c|}{Very Buggy} & \multicolumn{1}{c|}{All} & \multicolumn{1}{c|}{Buggy} & \multicolumn{1}{c|}{Very Buggy} \\ \hline
AspectJ & 593  & 4,439 & 2,281 & 602 & 2,394                           & 1,301                           & 1,350,239                & 355,361                    & 163,553                         \\ \hline
Tomcat  & 1,056 & 2,682 & 1,041 & 437 & 2,571                           & 1,966                           & 2,829,621                & 1,096,726                  & 459,506                         \\ \hline
SWT     & 4,151 & 2,249 & 1,181 & 657 & 8,281                           & 7,757                           & 9,327,318                & 4,894,050                  & 2,719,450                       \\ \hline
Eclipse & 5,262 & 4,614 & 2,868 & 1,581 & 10,907                          & 9,620                           & 33,206,245               & 24,267,961                 & 8,309,602                       \\ \hline
JDT     & 6,269 & 11,273 & 4,610 & 1,817 & 16,310                          & 11,940                          & 70,654,127               & 28,885,357                 & 11,378,833                      \\ \hline
\end{tabular}
}
\end{center}
\end{table*}

\begin{figure}
    \centering
    \includegraphics[width=0.5\textwidth]{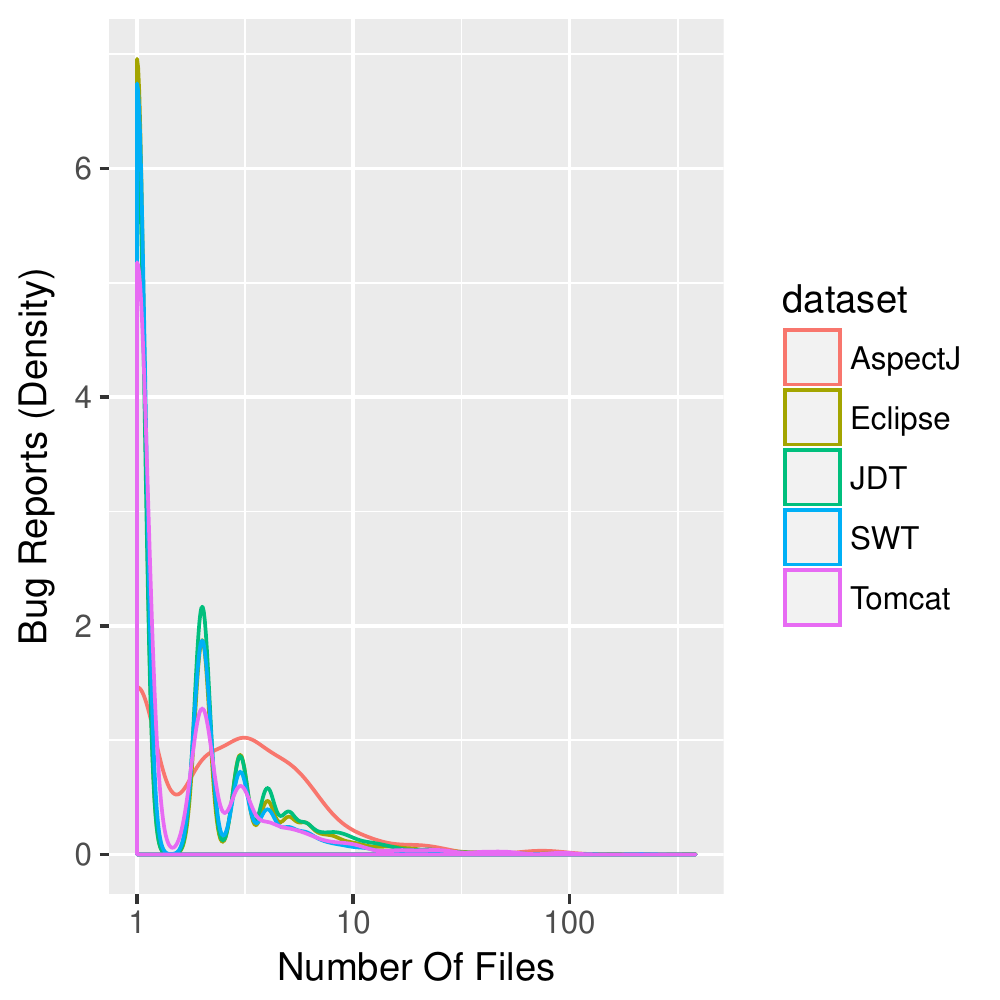}
    \caption{Analysis of linked records for all projects: distribution of the number of files changed per bug report}
    \label{fig:file_distr}
\end{figure}

\begin{figure}
    \centering
    \includegraphics[width=0.5\textwidth]{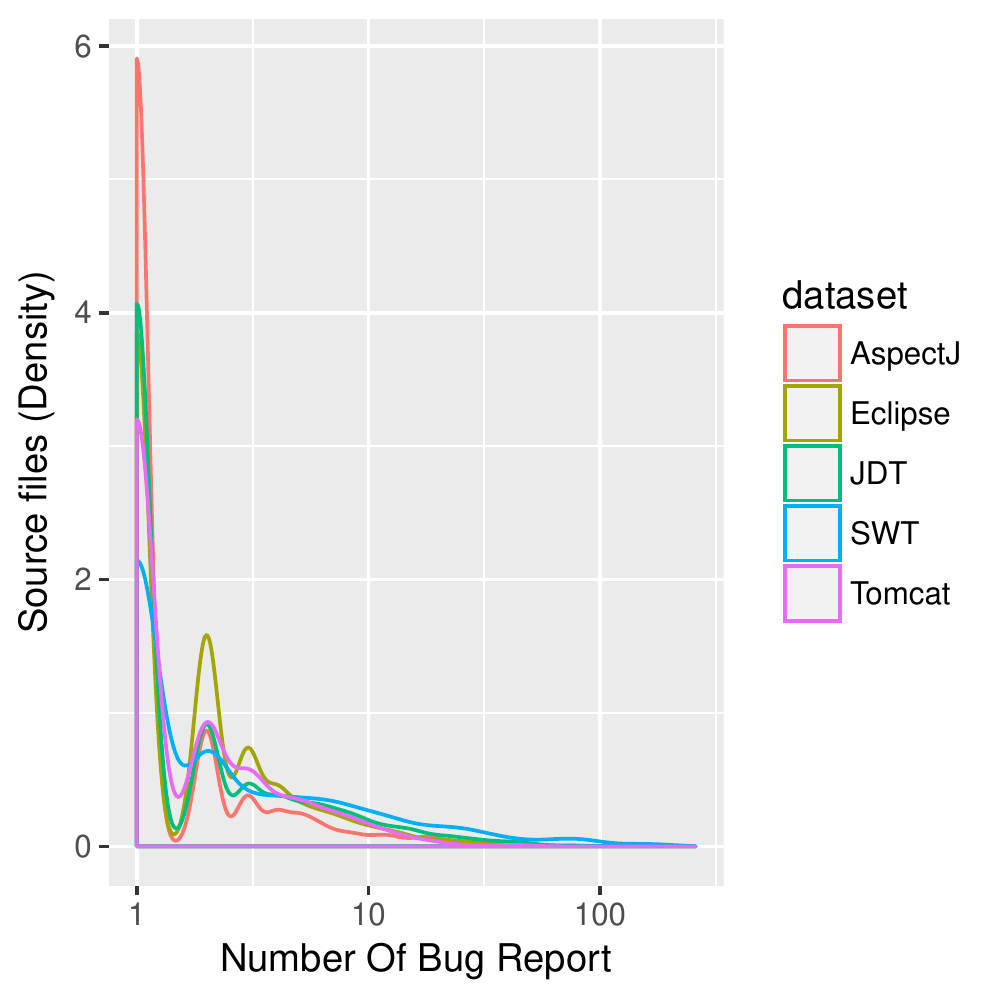}
    \caption{Analysis of linked records for all projects: distribution of the number of bug reports associated with a file}
    \label{fig:bug_distr}
\end{figure}

\subsection{Data Preprocessing}\label{sec:dat_prepr}
After we extract the data, we preprocess the corpus of bug reports (which is obtained from combining the summary and the description fields of the bug report) and the corpus of the source files by first removing punctuation symbols and numbers. Then, we remove Java-related keywords, comments, and package names from the source code corpus. Next, we apply Camel-case split and Porter stemming \cite{porter1980algorithm} on the bug report and source code corpus. Finally, we convert all the words/tokens to lowercase.

We then replicate data preprocessing scheme of~\cite{Lam-2015,Guo:2017:SES:3097368.3097370}, to be in line with the state-of-the-art models\footnote{Note that this scheme allows a look ahead into the future for non-linked files, which may make predictions overly-optimistic~\cite{tu2018}. In the spirit of replication study, we stick to this scheme. As we will see in the following sections, even these overly-optimistic predictions are not stellar.}. Next, we build the traceability matrix. We create an $m \times n$ Cartesian product of the list of all source files (both buggy and non-buggy files) and bug reports. I.e., if we have  $B$ bug reports and $S$ source files, we end up with a total of $B \times S$ records. We identify the linked or the relevant records $L$ (\textit{positive}) in the $B\times S$ records, based on the bug-file mapping. The remaining records in the traceability matrix, after removing the linked records will be the non-linked records, deemed $Z$ (\textit{negative}). Thus, the number of non-linked records in the traceability matrix will be $Z = (B \times S) - L$. Table \ref{tab::1} shows the linked and non-linked records statistics for each project. After this step, we merge the corpus of each bug report and its corresponding source file in the traceability matrix. Table~\ref{tab::1} shows that the datasets for all three variations of source files are highly  imbalanced, with  $L \ll Z$; the imbalance as well as the search space decrease  from All to Buggy to Very Buggy variation.

Next, we analyze the linked records in our traceability matrix. Figure \ref{fig:file_distr} shows the number of source files changed across the bug reports for all the projects in this study. As observed from the density plot, $\approx60\%$ of bug reports have one source file mapped to the bug report, $\approx18\%$ -- two files, $\approx8\%$ -- three files, etc. Rarely, namely in $\approx 0.06\%$ of the cases, more than 100 files are mapped to a bug. This shows that in order to resolve  $\approx 40\%$ of bug report, a developer has to fix code in more than one source file, which is not uncommon~\cite{li2011,khatri2011measuring}. It is also known that not all files in a bug-fixing commit may be directly related to that bug report~\cite{mills2018bug}. However, the  authors of the original study took all files in a bug-fixing commit. Therefore, we also took them all to follow the replication process.

Figure \ref{fig:bug_distr} shows a density plot of the number of bug reports associated with source files for all the projects. This density plot shows that $\approx 56\%$ of files are associated with a single bug, $\approx 17\%$ -- with two bugs, $\approx 8\%$ -- with three bugs, etc.  There also exists a small percentage ($\approx 0.07\%$) of outlier files associated with more than a 100 bugs. Note that this does not necessarily imply that these files are error-prone: they may not be related to an actual fix~\cite{mills2018bug}, as discussed above. Using the same rationale as in the previous paragraph, we retain all of the files.

\subsection{Experimental Setup}
In this section, we discuss the experiments performed in our study on the five open source datasets along with the exact setup used to train the CNN and Simple Logistic models.

\subsubsection{CNN Model}\label{sec:cnn}
The CNN model is implemented using the Keras deep learning library v.2.1.2 \cite{FAQKeras21:online} with TensorFlow v.1.4.1 \cite{tf:online} as the back-end. I/O and other non-compute intensive operations were performed on a CPU and the compute-intensive operations were carried out on 2 - 3 GPUs (depending on the availability of the  resources in a high performance compute cluster).

For each dataset, we first set aside 10\% of the data as test set and the remaining 90\% are be used for training and validation. The 90\% data are split into 10 folds. Then 10-fold cross-validation is performed. We balance the training data by applying higher class weights to the smaller class (i.e., positive class) which will penalize the model for every misclassificaltion of a positive sample (linked record). After each epoch, the trained model is tested on the imbalanced validation set and its cross entropy loss is recorded. At this point, we also compare the validation loss of the model with the previously recorded validation losses. If at a particular epoch the validation loss is lower than its previous values, then we save the weights of the model and use the same model to do the prediction on the 10\% test set.  
After all the epochs for a fold, the best model is obtained, i.e., the one with the lowest validation loss will be the final model that is used to make predictions on the imbalanced test set. The performance of the model is calculated based on the predictions made by the model on the test set. We repeat the same process for all the folds. The final performance of the model is the average of the performance of the model in each fold. 

Table \ref{tab::4} shows the hyper parameters used for training the model on all the datasets. Note that we are training the models on Eclipse and JDT corpuses for 5 epochs, on SWT and Tomcat -- for 10 epochs, and on AspectJ -- for 50 epochs. The reason for this partitioning lies in the correlation between the size of the text corpus and the amount of time needed to train the model. While the performance of the model may degrade with the decrease of the number of epochs, the duration to train the model may make it impractical, as we will further discuss in Section~\ref{sec:rq3}.

\begin{table}[t]
\caption{Hyperparameters for models}
    \centering
    \begin{tabular}{|c|c|}\hline
Parameter & Value  \\\hline
\multicolumn{2}{|c|}{CNN Model}\\\hline
Input Encoding & One-Hot \\\hline
Filter Size & $[2,3,4,5]$ \\\hline
Number Of Filters & $100$ \\\hline
Number Of Hidden Units & $100$ \\\hline
Activation Function & ReLU \\\hline
Drop-out Rate & $0.5$\\\hline
Batch Size & $16,64$ \\\hline
Optimiser & Adam \\\hline
Learning Rate & $10^{-4}$ \\\hline
Epochs & $5,10,50$ \\\hline
\multicolumn{2}{|c|}{Simple Logistic Model}\\\hline
Fixed Iterations & $0$ \\\hline
Max Boosting Iterations & $500$ \\\hline
Weight Trimming & $0.0, 0.1$ \\\hline
\end{tabular}
\label{tab::4}
\end{table}

\subsubsection{Simple Logistic Model}
We use Weka v.3.8.2 \cite{Runningf92:online} command-line interface to run the Simple Logistic models on all our datasets. The experimental set up is the same as that used to train the CNN models. In this case, we first convert the string attributes into numeric attributes representing the word occurrence information from the text contained in the strings using Weka's \textit{StringToWordVector} \cite{StringTo95:online}. Next, we apply a \textit{Class Balancer} \cite{ClassBal40:online} to the training data, to reweight the instances in the data so that each class has the same total weight. After this step, we apply the \textit{Simple Logistic} \cite{SimpleLo28:online} models on the dataset with 10-fold cross validation. The hyperparameters used to train the Simple Logistic models are given in Table \ref{tab::4}.

\subsection{Results \& Discussion}
We discuss answers to each of our research questions in the following sections.

During this discussion we will analyze the results of our experiments reported in Tables~\ref{tab::5} and \ref{tab::8}. The dash `-' in Tables \ref{tab::5} and \ref{tab::8} denotes experiments where 36GB of GPU RAM was insufficient to carry the computations. The star `*'in Tables \ref{tab::5} and \ref{tab::8} indicates that the computations for a given experiment could not be carried out, due to the lack of resources on a high performance computing cluster. We also could not get the resources on the compute cluster to process SWT and Eclipse `All Files' datasets on the cluster, so we leveraged the resources of our local server without GPUs (hence the zero GPU count in Table~\ref{tab::5}), but with 28 hyper-threading Intel Xeon CPU cores. Thus, the timing and resource utilization of these experiments should not be compared with those of the remaining CNN calibrations, due to different hardware setup. The calibration of Simple Logistic model for SWT `Buggy Files' corpus timed out after 25 days due to quota on  resources of the compute cluster. However, even with missing data points, we can extrapolate our findings based on the remaining computations, as discussed below.

\subsubsection{RQ1: Minimizing the Lexical Gap Between Bug Reports And Source Files}
The key drawback of the state-of-the-art bug localization approaches is that, they are unable to minimize the lexical gap \cite{bettenburg2008makes, Nguyen-2011, Ye:2014:LRR:2635868.2635874} between the natural language corpus of the bug reports and the technical corpus of the source code. Some attempts have been made in the past to overcome this drawback.  Ye et al.\cite{Ye:2014:LRR:2635868.2635874} used additional corpus from the documentation of the APIs used in source files. This approach did not help as the documentation of APIs does not have the buggy information specific to a project, rather they contain information regarding more general tasks. Kim et al. \cite{Kim} used the names of the previously fixed source files as classification labels on the bug reports instead of the actual source file content. 
In recent years some researchers have proposed deep-learning-based models to localize bugs.
Lam et al. \cite{Lam-2015, Lam-2017} and Huo et al. \cite{Huo-2016, Huo-2017} proposed deep learning models based on RBM, CNN, LSTM. Their work proved that, unlike the other state-of-the-art traditional models, the deep learning models are able to minimize the lexical gap between bug reports and source files. They concluded that DNNs are able to do this by learning to link high-level, abstract concepts between bugs reports and source code files.

As mentioned earlier, a simple CNN with little hyper-parameter tuning has been shown to perform significantly well for classifying text \cite{kim2014convolutional}. CNNs construct higher-order features  from lower-order features and preserve the local information about locality. They lose the global information about locality due to Local Invariance. Nevertheless, this does not affect the performance of the model when performing sentiment analysis or text classification, as in these cases the ordering of words is not of importance at the document level. This means that, CNNs are good at identifying the polarity of a sentence but cannot encode long-range dependencies in a sentence.

As discussed earlier, bug localization is a Learning-To-Rank IR problem which can be solved using the Pairwise approach, where each sentence is a combination of bug reports and source files and the task is to identify the positive (linked) and negative (non-linked) records. Hence a CNN model, which has been widely used in the past for learning the polarity of a sentence \cite{kim2014convolutional} could be a potential model that can correctly classify the corpus related to bug report and source code files, by learning to relate the natural language or domain-related terms/tokens in bug reports and different code tokens in source files. 

\medskip\fbox{\begin{minipage}{0.95 \textwidth}
A deep-learning-based model like CNN could be a potential approach that could eliminate or minimize the lexical gap between bug reports and source code files. 
\end{minipage}}

\begin{table}[t]
\caption{Models' resource consumption and training time}
\centering
    \begin{tabular}{|l|r|r|r|r|r|r|} 
    \hline
        \multicolumn{1}{|c|}{Dataset}  & \multicolumn{4}{|c|}{CNN} & \multicolumn{2}{|c|}{Simple Logistic}\\\hline
          & Memory (GB) & GPUs &  Time (days) & Time per epoch (days) & Memory (GB) & Time (days) \\\hline
         \multicolumn{7}{|c|}{`All Files'}\\\hline
         AspectJ & 40 & 2 & 21 & 0.4 & 250 & 17\\\hline
         Tomcat & 80 & 3 & 70 & 7 & 550 & * \\\hline
         SWT & 145 & 0 & 150 & 15 & 3000 & * \\\hline
         Eclipse & 170 & 0 & 150 & 30 & 3000 & * \\\hline
         JDT & - & - & - & - & & *\\\hline
         \multicolumn{7}{|c|}{`Buggy Files'}\\\hline
         AspectJ & 10 & 1 & 7 & 0.1 & 50 & 4\\\hline
         Tomcat & 80 & 3 & 24 & 2.4 & 80 & 13\\\hline
         SWT & 120 & 3 & 110 & 11 & 2500 & $>$25 \\\hline
         Eclipse & 120 & 3 & 110 & 22 & 2500 & * \\\hline
         JDT & - & - & - & - & & *\\\hline
         \multicolumn{7}{|c|}{`Very Buggy Files'}\\\hline
         AspectJ & 10 & 1 & 3 & 0.1 & 50 & 2\\\hline
         Tomcat & 100 & 3 & 6 & 0.6 & 300 & 4\\\hline
         SWT & 120 & 3 & 60 & 6 & 800 & * \\\hline
         Eclipse & 120 & 3 & 50 & 10 & 2500 & * \\\hline
         JDT & 120 & 3 & 80 & 16 & 2500 & * \\\hline
\end{tabular}
\label{tab::5}
\end{table}

\begin{table}[t]
        \caption{Models' performance}
      \centering
        \begin{tabular}{|l|r|r|r|r|r|r|r|r|}
        \hline
         \multicolumn{1}{|c|}{Dataset}  & \multicolumn{4}{|c|}{CNN Model} & \multicolumn{4}{|c|}{Simple Logistic Model}\\\hline
  & AUC & MAP & MRR & Top-5 & AUC & MAP & MRR & Top-5 \\\hline
  \multicolumn{9}{|c|}{`All Files'}\\\hline
AspectJ & 0.9 & 0.25 & 0.27 & 44\% & 0.9 & 0.3 & 0.33 & 52\% \\\hline
Tomcat & 0.84 & 0.16 & 0.17 & 23\% & *  & *  & * & *  \\\hline
SWT & 0.86 & 0.24 & 0.26 & 38\% & * & * & *  &  *\\\hline
Eclipse & 0.6 & 0.04 & 0.04 & 5\% & * & *  & * & * \\\hline
JDT & - & - & - & - & * & * & * & *\\\hline
\multicolumn{9}{|c|}{`Buggy Files'}\\\hline
AspectJ & 0.86 & 0.34 & 0.38 & 61\% & 0.84 & 0.43 & 0.47 & 63\% \\\hline
Tomcat & 0.84 & 0.29 & 0.32 & 40\% & 0.77 & 0.05 & 0.05 & 5\%  \\\hline
SWT & 0.87 & 0.3 & 0.3 & 51\% & * & * & * & * \\\hline
Eclipse & 0.72 & 0.13 & 0.14 & 18\% & * & *  & * & * \\\hline
JDT & - & - & - & - & * & * & * & * \\\hline
\multicolumn{9}{|c|}{`Very Buggy Files'}\\\hline
AspectJ & 0.83 & 0.44 & 0.47  & 64\% & 0.79 & 0.16 & 0.17 & 24\% \\\hline
Tomcat & 0.81 & 0.35 & 0.37 & 52\% & 0.66 & 0.25 & 0.26 & 36\%  \\\hline
SWT & 0.82 & 0.3 & 0.3 & 51\% & * & * & *  & * \\\hline
Eclipse & 0.75 & 0.16 & 0.17 & 23\% & * & *  & * & * \\\hline
JDT & 0.67 & 0.11 & 0.12 & 17\% & * & * & *  & *\\\hline
\end{tabular}
\label{tab::8}
\end{table}

\begin{table}
\caption{Comparison of HyLoc with other models \cite{Lam-2015}}
\centering
\begin{tabular}[t]{|c|c|c|c|c|}\hline
Dataset & Model & Top-5 & $MRR$ & $MAP$  \\\hline

AspectJ & HyLoc & 71.2\% & 0.52 & 0.32 \\
 & LR & 45.5\% & 0.33 & 0.25 \\
 & BL & 47.7\% & 0.32 & 0.22 \\
 & NB & 16\% & 0.1 & 0.07 \\\hline

Tomcat & HyLoc & 72.9\% & 0.6 & 0.52 \\
 & LR & 66.5\% & 0.55 & 0.49 \\
 & BL & 61.8\% & 0.48 & 0.43 \\
 & NB & 9\% & 0.08 & 0.07 \\\hline
 
SWT & HyLoc & 69.0\% & 0.45 & 0.37 \\
 & LR & 58.2\% & 0.41 & 0.36 \\
 & BL & 38.3\% & 0.28 & 0.25 \\
 & NB & 19\% & 0.14 & 0.11 \\\hline

Eclipse & HyLoc & 70.5\% & 0.51 & 0.41 \\
 & LR & 60.1\% & 0.47 & 0.40 \\
 & BL & 49.3\% & 0.37 & 0.31 \\
 & NB & 10.6\% & 0.07 & 0.06 \\\hline

JDT & HyLoc & 65\% & 0.45 &  0.34 \\
 & LR & 55.2\% & 0.42 & 0.34\\
 & BL & 40.2\% & 0.30 & 0.23 \\
 & NB & 15\% & 0.11 & 0.08 \\\hline
\end{tabular}
\label{tab::11}
\end{table}

\begin{table}
\caption{Comparison of DNN-based models with other models \cite{Huo-2017}}
\centering
\begin{tabular}[t]{|c|c|c|c|}\hline
Dataset & Model & Top-10 & $MAP$  \\\hline

AspectJ
 & BL & 62.2\% & 0.41 \\
 & BLUiR & 65.9\% & 0.43 \\
 & AmaLgam & 69.3\% & 0.42 \\
 & CNN & 77.3\% & 0.51  \\
 & NP-CNN & 83.6\% & 0.54  \\
 & LSTM & 79.2\% & 0.51 \\
 & LSTM+ & 85.5\% & 0.54 \\
 & LS-CNN & 86.9\% & 0.56 \\\hline
 
Eclipse
 & BL & 72.7\% & 0.42 \\
 & BLUiR & 75.4\% & 0.44  \\
 & AmaLgam & 77.3\% & 0.44  \\
 & CNN & 82.1\% & 0.49\\
 & NP-CNN & 87.2\% & 0.54 \\
 & LSTM & 86.9\% & 0.52\\
 & LSTM+ & 87.2\% & 0.54 \\
 & LS-CNN & 89.5\% & 0.56  \\\hline
 
JDT
 & BL & 70.3\% & 0.44 \\
 & BLUiR & 74.5\% & 0.43 \\
 & AmaLgam & 75.7\% & 0.44 \\
 & CNN & 85.9\% & 0.51 \\
 & NP-CNN & 88.2\% & 0.53\\
 & LSTM & 86.8\% & 0.52 \\
 & LSTM+ & 88.3\% & 0.54 \\
 & LS-CNN & 91.7\% & 0.58 \\\hline
\end{tabular}
\label{tab::12}
\end{table}

\subsubsection{RQ2: Effectiveness}
To assess practical relevance of the models we analyze our results (shown in Tables~\ref{tab::5} and \ref{tab::8}) with the results of other state-of-the-art models assessed in \cite{Lam-2015,Huo-2017} for which the results are shown in Tables~\ref{tab::11} and \ref{tab::12}. These models include DNNs (namely,  \textit{HyLoc} \cite{Lam-2015}, \textit{NP-CNN} \cite{Huo-2016}, and \textit{LS-CNN} \cite{Huo-2017}) as well as state of the art classic ML models (namely, \textit{BugLocator} (BL) \cite{Zhou}, 
\textit{Two-Phase} (NB) \cite{Kim}, \textit{Learning-To-Rank} (LR) \cite{Ye:2014:LRR:2635868.2635874}, \textit{BLUiR} (Bug Localization Using Information Retrieval) \cite{Saha}, \textit{AmaLgam} \cite{Wang-2014}).

Direct comparison of the performance of the models is challenging, as every group of authors uses different data preparation techniques, hardware, and software. Our data preparation process resembles that of \cite{Huo-2016,Huo-2017} (the latter paper is an extension of the former one). Performance of the CNN model is assessed in \cite{Huo-2017}, but their dataset has a different number of bugs and files.

When comparing these two sets of tables, we notice that the performance of the CNN model in our experiments is inferior to those of~\cite{Huo-2017}. This can be attributed to the fact that they experimented on the reduced set of source files (as we found out based on direct communication with the authors). This setup may be acceptable to compare the performance of a set of models, but it may be misleading the practitioners, as the resulting values of the metrics do not reflect the performance of the model on a complete set of files, which is what practitioners are interested in. The actual effectiveness of any bug localization model can only be determined when the model is trained on the entire set of source code files available in the project repository. 

Let us assess the effectiveness of the model based on survey~\cite{Kochhar:2016:PEA:2931037.2931051} summarized in Section~\ref{sec:Background}.

\textbf{Granularity}: Since all of the models mentioned above operate on the file-level, they satisfy only  26\% of the practitioners in terms of granularity.

\textbf{Success Criteria}: Huo and Li~\cite{Huo-2017} report performance of the models using Top-10 metric and do not provide Top-5 values; based on~\cite{Kochhar:2016:PEA:2931037.2931051}, only 17\% of practitioners would be satisfied with Top-10 results. Lam et al. as well our experiments do provide Top-5 metrics. Let us see if practitioners would find these results trustworthy. 

\textbf{Trustworthiness}: Based on~\cite{Kochhar:2016:PEA:2931037.2931051} the value of Top-5 metric is directly correlated with the measure of trustworthiness: that is Top-5 = 70\%  translates into 70\% trustworthiness. Based on Tables~\ref{tab::8} and \ref{tab::11}, the trustworthiness varies between 17\% and 73\%, depending on the model. 

Given that Table \ref{tab::12} reports Top-10 values, it is more challenging to assess these results. We conjecture that the values of Top-5 (reported in Table \ref{tab::12}) will be lower than Top-10 and would probably be even lower if the models were tested on the full set of files.

Overall, while comparing ``goodness-of-fit'' of the DNN-based and classic-ML-based models in Tables~\ref{tab::8}, \ref{tab::12}, and \ref{tab::12}, we see that the former outperform the latter most of the time (with two exceptions in \ref{tab::8}).

\textbf{Scalability}: Conceptually, the models in our study can process  code bases as large as  JDT (which has more than 1,000,000 LOC), satisfying up to 90\% of practitioners. However, the required amount of time or compute resources needed to train a model may be excessive. As Table~\ref{tab::5} shows, in the worst case scenario, we may require terabytes of RAM, multiple GPUs, and months of compute time, making training impractical. Huo and Li~\cite{Huo-2017} do not provide their timing data, but Lam et al.~\cite{Lam-2015} do share these data. The HyLoc model~\cite{Lam-2015} can be trained in less than 1 day on a 32 core Xeon computer with 126GB of RAM. This is an excellent performance which can probably be attributed to the fact that Huo and Li extract  salient features from the text corpuses, reducing the dimensionality of the input data. This may be more challenging for practitioners, as they will need to find an expert capable of configuring and running such NLP dimensionality reduction techniques, which is more challenging than running off-the-shelf models implemented in Keras or Weka.

\textbf{Efficiency}: as stated in~\cite{Kochhar:2016:PEA:2931037.2931051}, a trained model should return a prediction in less than a minute. Our predictions take approximately 4 seconds per 1000 files (easily satisfying the constraint). Huo and Li \cite{Huo-2017} do not report their timing. Lam et al. HyLoc  model takes ``a few minutes''\cite{Lam-2015}. Given that prediction involves evaluation of each bug-file pair individually and independently, this embarrassingly parallel problem can be easily parallelized reaching the desired threshold of one minute (assuming that a practitioner is willing to pay for compute resources).

\medskip\fbox{\begin{minipage}{0.95 \textwidth}
\begin{itemize}
    \item Researchers should  use the entire set of source files from the repository when assessing practicality of a particular bug localization approach.
    
    \item DNN-based models outperform classic ML-based models most of the time. Yet, researchers should verify if their models meet the criteria stated by software practitioners in \cite{Kochhar:2016:PEA:2931037.2931051}.
    
    \item Software developers should be cautious while using the current bug localization models, as their performance varies.
\end{itemize}\end{minipage}}

\subsubsection{RQ3: CNN vs Simple Logistic Models}\label{sec:rq3}
We now compare a non-linear DNN (CNN model) and a traditional ML (Simple Logistic model) in terms of performance, training time and memory.

\textbf{Performance}: While we could not assess performance of the model for all datasets due to lack of compute resources (as shown in Tables~\ref{tab::5} and ~\ref{tab::8}), we do have complete results for AspectJ and partial -- for Tomcat. In case of AspectJ, the Simple Logistic model outperforms the CNN model in terms of $MAP$, $MRR$, and Top-5 Rank, when considering all the source files in the dataset.The same trend continues even when we train the model on only the buggy source files in the dataset. This trend is reversed when we further reduce the buggy source files. In this case, the CNN model performs significantly better than the Simple Logistic model. In the case of Tomcat, the Simple Logistic model outperforms CNN for `Buggy Files' and `Very Buggy Files'; we do not have results for `All files', which can be explained by the smaller number of epochs used in our study (10 epochs for Tomcat vs 50 epochs for AspectJ). 
    
\textbf{Training Time}: The training time for each dataset (given in Table~\ref{tab::5}), is influenced by the number of GPUs used. If we compare the training times across projects, per GPU, then CNN takes between 3 and 42 days for the smallest project and between 150 and 330 days for larger projects like Eclipse. For the Simple Logistic model and small datasets, the training time is comparable magnitude-wise to the overall training time of the CNN model. 

However, we have to keep in mind that the number of epochs for large datasets was smaller than that for small datasets (as discussed in Section~\ref{sec:cnn}). Thus, we need to look at the per-epoch training time in Table~\ref{tab::5}. In this case we can see that Simple Logistics model outperforms CNN, if we keep the number of epochs at 50: e.g., calibration of CNN for Tomcat `Buggy Files` dataset would take 120 ($=$ 2.4 $\times$ 50) days.
    
\textbf{Memory}: To train, the CNN models need from 10GB to 150GB of memory, while the Simple Logistic models require  50GB to 3000GB of memory. Economically, CNN is cheaper to train as, at the time of writing, the cost of three GPUs is cheaper than 3000GB of RAM.

\medskip\fbox{\begin{minipage}{0.95 \textwidth}
\begin{itemize}
    \item Obtaining the computation resources needed to train the deep-learning-based or the traditional ML-based bug localization models,  is both expensive and challenging for the mainstream software practitioners.
    
    \item The CNN model outperforms the Simple Logistic model in most of the cases, but they have high training time. Simple Logistic models are faster but need a lot of memory to train (and, thus, are less feasible economically).
\end{itemize}
\end{minipage}}

\subsubsection{RQ4: Generalizability}
From Table~\ref{tab::8}, we can observe that for all the three variations of source files, the $AUC$ for the CNN models for all the datasets is between 0.6 and 0.9 (yielding some separation of classes).
The higher the $MAP$, the better is the performance of the IR model. As most of the values are around 0.8, we can conclude that the CNN can successfully classify different bug localization dataset.  The trend in performance of the $MAP$ across all the projects, for all the three variations of the source files, shows that the CNN models perform poorly for larger projects like Eclipse and JDT. This could be attributed to the fact that, for the larger projects, the datasets are more imbalanced and the search space is larger as compared to the smaller projects (see Table~\ref{tab::1} for details). This makes it even more difficult for the models to identify or learn patterns in the few sets of linked records are present in the dataset. The $MRR$ and  $MAP$ metrics trends are similar.

\medskip\fbox{\begin{minipage}{0.95 \textwidth}
The performance of the CNN models is dependent on the size of the dataset under study, as it performs significantly better on smaller datasets than on the larger datasets for all the three variations of source files.
\end{minipage}}

\begin{itemize}
    \item Obtaining the computation resources needed to train the deep-learning-based or the traditional ML-based bug localization models,  is both expensive and challenging for the mainstream software practitioners.
    
    \item The CNN model outperforms the Simple Logistic model in most of the cases, but they have high training time. Simple Logistic models are faster but need a lot of memory to train (and, thus, are less feasible economically).
\end{itemize}

\subsubsection{RQ5: Effect of Varying Buggy Files}
We now examine the effect of varying the buggy source files in the datasets. When reducing the number of source files, the performance of the CNN models improves, which can observed in Table~\ref{tab::8}. For all the projects, i.e., AspectJ, Tomcat, SWT, and Eclipse, the $MAP$ scores increase from 0.25 to 0.44, 0.16 to 0.35, 0.26 to 0.3, 0.04 to 0.17, respectively. The same observation can also be made for $MRR$ and Top-5 Rank metrics. We conjecture that this improvement may be because reducing the number of files is reducing the search space, i.e., reserving more relevant files actually enables the model to easily identify the linked or the relevant files in the dataset. As for the Simple Logistic models, the effect of varying source files is not very clear, as in some cases it tends to improve the performance of the model, whereas in some cases it does not.

\medskip\fbox{\begin{minipage}{0.95 \textwidth}
Considering `Very Buggy' files improves the performance of the CNN models. An important point to note from the above observation is that, most of the recent deep-learning-based models~\cite{Huo-2016, Lam-2015, Lam-2017, Huo-2017} consider only a subset of files  when assessing  the performance of the DNN models. This leads to overly-optimistic performance assessment results from the practitioners' perspective. 
\end{minipage}}

\subsection{Threats to Validity}\label{sec:threats}
In this section we discuss the threats to validity, classified as per \cite{wohlin2012experimentation,yin2009case}.

\textit{Internal Validity}:
We mitigate the bias in our study, by reusing the bug reports dataset that has been used in prior studies \cite{Huo-2016, Lam-2015, Lam-2017, Huo-2017, Ye:2014:LRR:2635868.2635874, Zhou}. To reduce the experimental errors, we have carefully checked
our implementation to the best of our abilities.

\textit{Construct Validity}:
Construct validity relates to the applicability of the set of evaluation metrics used in this study. The metrics like $MAP$, $MRR$ and Top-$k$ Rank are well-known information retrieval metrics and have been used before to evaluate many past bug localization approaches \cite{Rao-2011, Saha, sisman2012incorporating, ZhouZhang}. Thus, we believe there is little threat to construct validity. Another threat comes from the data preprocessing scheme as per~\cite{Lam-2015,Guo:2017:SES:3097368.3097370}, which  has deficiencies discussed in Section~\ref{sec:dat_prepr}. However, we retain the scheme to preserve the spirit of replication and align this work with the prior art. These deficiencies may lead to overly-optimistic predictions; however, even these inflated results may not be adequate for the practitioners, as was discussed in the previous sections.

\textit{Conclusion Validity}:
We made sure the model is generalizable by avoiding overfitting. In order to prevent overfitting, we have employed the dropout technique, 10-fold cross validation, regularization and early stopping. Also, the risk of insufficient generalization is reduced by evaluating the models on five open source projects.

\textit{External Validity}:
Software engineering studies suffer from the variability of the real world, and the \textit{generalization} problem cannot be solved completely  \cite{26c990771bd645428c33ea107259ceb5}. Although we have used five open source projects in this study, our empirical evaluation may not be generalizable to other open source projects or industrial projects. The goal of the study was not to build a new model, but to experiment on the recently proposed deep-learning-based bug localization models and examine their relevance in research and industry. The same empirical examination can also be applied to other software products with well-designed and controlled experiments.

\section{Conclusion}
In this study we examine the effectiveness of deep-learning-based bug localization models from the practitioners' perspective and compare them with traditional ML models. To achieve this, we assess performance of multiple models on five open source bug localization datasets. We assess the performance of two models directly and use readily available statistics~\cite{Lam-2015,Huo-2017} for the remaining models.  Finally, we evaluate the models against the expectations of  software practitioners (reported in~\cite{Kochhar:2016:PEA:2931037.2931051}) and discuss the drawbacks of the models. Our study includes the following key findings:

\begin{itemize}
    \item The deep-learning-based models perform well, compared to the traditional ML models on bug localization data in most cases, but have high training time and need large amount of hardware resources, such as GPUs and memory. 
    
    \item It is both expensive and challenging for the mainstream software practitioners to be able to use these models.
    
    \item It is challenging to compare the performance of different models when no standard approach for data preparation exists. Moreover, reductions in the sets of files used for testing the model may lead to overly optimistic performance evaluation from a practical standpoint. 
    
    \item Most researchers consider only the IR metrics to evaluate the performance of their proposed approach or model to localize bugs. In addition to this, the models should be examined to verify if they meet the expectations of a software practitioner. 

\end{itemize}

This work is of interest to software practitioners, as it provides evidence that the practitioners should be cautious while using the current state of the art models. It is also of interest to researchers as it highlights the need for standardization of performance benchmarks to ensure realistic assessment of bug localization models.

\printbibliography
\end{document}